\documentclass[12pt]{article}
\usepackage{graphicx}
\usepackage{epsfig}
\begin{document}
\topskip 1cm
\renewcommand{\thefootnote}{\fnsymbol{footnote}}  
\begin{titlepage}
\rightline{ \large{ \bf Nov 2003} }
\begin{center}
{\Large\bf Numerical evaluation of the general massive }\\ 
{\Large\bf 2-loop 4-denominator self-mass master integral }\\
 {\Large\bf from differential equations.}
\footnote{Supported in part by the EC network EURIDICE, 
contract HPRN-CT-2002-00311 and by 
Polish State Committee for Scientific Research
(KBN) under contract No. 2 P03B 017 24.}
\\

\vspace{2.cm}
{\large {\bf
M.~Caffo$^{ab}$, 
H.~Czy{\.z}\ $^{c}$, 
A.~Grzeli{\'n}ska$^{c}$ }
and   
{\bf E.~Remiddi$^{ba}$ } }\\

\begin{itemize}
\item[$^a$]
             {\sl INFN, Sezione di Bologna, I-40126 Bologna, Italy }
\item[$^b$] 
             {\sl Dipartimento di Fisica, Universit\`a di Bologna, 
             I-40126 Bologna, Italy }
\item[$^c$] 
             {\sl Institute of Physics, University of Silesia, 
             PL-40007 Katowice, Poland }

\end{itemize}
\end{center}

\noindent
e-mail: {\tt caffo@bo.infn.it \\ 
\hspace*{1.3cm} czyz@us.edu.pl \\ 
\hspace*{1.3cm} grzel@joy.phys.us.edu.pl \\ 
\hspace*{1.3cm} remiddi@bo.infn.it \\ } 
\vspace{.5cm}
\begin{center}
\begin{abstract}
The differential equation in the external invariant $p^2$ satisfied by the 
master integral of the general massive 2-loop 4-denominator self-mass
diagram is exploited and the expansion of the master integral at $p^2=0$
is obtained analytically. 
The system composed by this differential equation with those of the 
master integrals related to the general massive 2-loop sunrise diagram
is numerically solved by the Runge-Kutta method in the complex $p^2$ plane.
A numerical method to obtain results for values of $p^2$ at and
close to thresholds and pseudo-thresholds is discussed in details. 
\end{abstract}
\end{center}
\scriptsize{ \noindent ------------------------------- 
    \\ } 
\vfill
\end{titlepage}

\pagestyle{plain} \pagenumbering{arabic} 
\newcommand{\Eq}[1]{Eq.(\ref{#1})} 
\newcommand{\labbel}[1]{\label{#1}} 
\newcommand{\cita}[1]{\cite{#1}} 
\def\Re{\hbox{Re~}} 
\def\Im{\hbox{Im~}} 
\newcommand{\Sm}{S(n,m_1^2,m_4^2,p^2)} 
\newcommand{\Rs}{R^2(m_1^2,m_2^2,m_3^2)} 
\newcommand{\RRp}{R^2(-p^2,m_1^2,m_4^2)} 
\newcommand{\V}{V(n,m_1^2,m_2^2,m_3^2)} 
\newcommand{\G}{G(n,m_1^2,m_2^2,m_3^2,m_4^2,p^2)} 
\newcommand{\F}[1]{F_#1(n,m_1^2,m_2^2,m_3^2,p^2)} 
\newcommand{\FF}[1]{F_#1(n,m_1^2,m_2^2,m_3^2,m_4^2,p^2)} 
\newcommand{\dnk}[1]{ d^nk_{#1} } 
\newcommand{\Fn}[2]{F_{#1}^{(#2)}(m_1^2,m_2^2,m_3^2,p^2)}
\newcommand{\FFn}[2]{F_{#1}^{(#2)}(m_1^2,m_2^2,m_3^2,m_4^2,p^2)}
\newcommand{\D}{D(m_1^2,m_2^2,m_3^2,p^2)} 
\def\Li2{\hbox{Li}_2} 
\def\LLL{L(m_1^2,m_2^2,m_3^2)} 
\def\a{\alpha} 
\def\app{{\left(\frac{\alpha}{\pi}\right)}} 
\newcommand{\e}{{\mathrm{e}}} 
\newcommand{\om}{\omega} 
\newcommand{\verso}[1]{ {\; \buildrel {n \to #1} \over{\longrightarrow}}\; } 

\section{Introduction.} 

Precise measurements of particles properties require that the 
corresponding theoretical calculations have to include 
up to (at least) two-loop radiative corrections. In this paper we 
investigate a fast and  flexible numerical method for their accurate 
evaluation.

A commonly used procedure in modern radiative correction 
calculations is to express the result as a combination of 
a limited number of Master Integrals (MI's), 
using the integration by parts identities \cite{TkaChet}. 
In this framework, it is necessary to obtain a precise determination 
of the MI's, even if the analytical values cannot be obtained due to 
the large number of different scales occurring in each of the MI's 
(internal masses and 
external momenta or Mandelstam variables), as it happens in electroweak
theory. 

The general massive 2-loop self-mass diagram has four MI's related to the
sunrise (3-denominator) diagram, only one independent MI related to the 
4-denominator diagram, and again only one new MI related to the 
5-denominator diagram \cite{Tarasov,MS,CCLR1,CCLR2}.

Several procedures for a precise numerical evaluation of the MI's have 
been and still are investigated, such as multiple expansions \cite{BBBS}, 
 numerical integration \cite{BBBS,BT,BBBB,GvdB,PT,GKP,ABT,Pass,PU}, or 
difference equations \cite{Laporta}.

Another method based on differential equations \cite{Kot,ER} was
proposed in \cite{CCR3}, where it was shown how to use Runge-Kutta
method \cite{PTVF} to solve numerically 
the system of linear differential equations
in the external invariant $p^2$ 
satisfied by sunrise master integrals.
\begin{figure}[!h]
\epsfbox[-40 40 140 180]{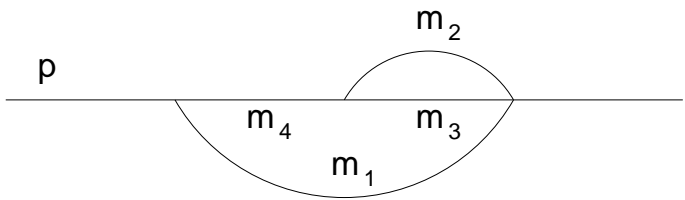}
\caption{The general massive 2-loop 4-denominator self-mass diagram.} 
\label{fig:f0}
\end{figure}
The method was extended to 2-loop 4-denominator and 5-denominator
cases in \cite{Martin}, where it was also suggested how to evaluate
numerically the MI's nearby thresholds and pseudo-thresholds.
We give in this paper an accurate implementation of that approach 
for the sunrise and the 4-denominator MI's.

In Section 2, the $(n-4)$ expansions of the MI's are constructed, in 
Section 3 initial conditions for the differential equations for 
the Master Integrals (or Master Differential Equations, MDE's) are discussed. 
Some results of the program, showing characteristic behaviour of the 
studied MI's are presented in Section 4.
In Section 5, a method for the numerical
evaluation of the MI's near thresholds and pseudo-thresholds is
discussed in detail, 
while Section 6 is devoted to comparisons with existing results.

Finally, in Section 7, our conclusions on the application of the method 
to present and further work are presented.

\section{The MDE and the expansion in $(n-4)$ of the MI} 

We use here the following definition of the MI related to the 
general massive 2-loop 4-denominator self-mass diagram, shown in 
Fig.{\ref{fig:f0}}, 
\begin{eqnarray} 
 \FF{4}
      &=& \frac{ \mu^{8-2n}}{((2\pi)^{n-2})^2 }  
         {\kern20pt} \int \dnk{1} \int \dnk{2} \nonumber \\  
      && {\kern-165pt} \frac{1} 
 {(k_1^2+m_1^2)\ [(p-k_1)^2+m_4^2]\ (k_2^2+m_2^2)\ [(p-k_1-k_2)^2+m_3^2]} \ ,
\labbel{MI} \end{eqnarray} 

\noindent
where integration is performed in $n-$dimensional Euclidean space.
\footnote{$\FF{4}$ of the present paper corresponds to $ C^2(n) \G$ 
of \cite{CCLR2}.}

Wherever necessary to avoid ambiguities, the usual imaginary displacements 
in the masses $m_i^2 \to m_i^2 -i \epsilon$, where $\epsilon$ is an 
infinitesimal positive number, are understood.  
The arbitrary mass scale $\mu$ accounts for the continuous value of the 
dimensions $n$.
In numerical calculations, we choose $ \mu = m_1+m_2+m_3 $,
one of the natural scales of the problem, corresponding to the 3-body 
threshold, while for simplicity in all analytic formulae we put $\mu =1$. 
To recover results for arbitrary $\mu$, one has to substitute 
$m_i \to {m_i}/{\mu}$ and $p^2 \to {p^2}/{\mu^2}$.

The master equation reads 
\footnote{Note the change in sign in the 3rd line of \Eq{ME}  
in comparison to \cite{CCLR2}, due to a misprint there.}
\begin{eqnarray} 
  && {\kern-30pt} \RRp \ p^2 \ \frac{\partial}{\partial p^2} \FF{4} 
  = \nonumber \\ 
  && \phantom{+} \frac{n-4}{2} \RRp \FF{4}                 \nonumber \\ 
  && - (n-3)\left[ (m_1^2+m_4^2)p^2 + (m_1^2-m_4^2)^2 \right] \FF{4} 
                                                           \nonumber \\ 
  && + (3p^2-m_1^2+m_4^2) m_1^2 \F{1}                      \nonumber \\ 
  && + (p^2-m_1^2+m_4^2) \Biggl[ \frac{3n-8}{2}  \F{0}     \nonumber \\ 
  && {\kern+30pt} + m_2^2 \F{2} + m_3^2 \F{3}              \nonumber \\ 
  && {\kern+30pt} - \frac{1}{2}(n-2) V(n,m_2^2,m_3^2,m_4^2) \Biggr] \ ,
\labbel{ME} \end{eqnarray} 
where 
\begin{eqnarray} 
 \RRp &=& p^4+m_1^4+m_4^4+2m_1^2p^2+2m_4^2p^2-2m_1^2m_4^2    \nonumber \\ 
      &=& [p^2+(m_1+m_4)^2] [p^2+(m_1-m_4)^2] \ .
\labbel{RRp} \end{eqnarray} 
The \( \F{j}, j=0,1,2,3 \) are the massive 2-loop sunrise 
self-mass master amplitudes, discussed in \cita{CCLR1} and numerically 
calculated in \cita{CCR3},
\begin{eqnarray} 
 \F{j}
      &=& \frac{ \mu^{8-2n}}{((2\pi)^{n-2})^2 } 
         {\kern20pt} \int \dnk{1} \int \dnk{2} \nonumber \\
      && {\kern-150pt} \frac{ 1 } 
           { (k_1^2+m_1^2)^{\alpha_1(j)} (k_2^2+m_2^2)^{\alpha_2(j)} 
             ( (p-k_1-k_2)^2+m_3^2 )^{\alpha_3(j)} } \ , \quad j=0,1,2,3 
\labbel{SMI} \end{eqnarray} 
where: for $j=0,\ i=1,2,3, \ \alpha_i(0)=1;\ $ for $j,i=1,2,3,\ $
\(\alpha_i(j)=1\), if \(i\ne j\); \(\alpha_i(j)=2\), for \(i = j\).
The function \( \V \), finally, is the massive 2-loop sunrise 
vacuum amplitude defined in \cita{CCLR1} (see also \cita{Ford,Tausk}) 
\begin{eqnarray} 
    \V 
      = \frac{ \mu^{8-2n}}{((2\pi)^{n-2})^2 } 
          \int \dnk{1} \int \dnk{2} 
      \frac{1} 
      {(k_1^2+m_1^2)\ (k_2^2+m_2^2)\ [(k_1+k_2)^2+m_3^2]} \ .
\labbel{SV} \end{eqnarray} 
In the following we also use the massive 1-loop self-mass 
\begin{equation}
 \Sm = \frac{\mu^{4-n}}{(2\pi)^{n-2}} \int d^nk \ 
    \frac{1} {(k^2+m_1^2)\ [(p-k)^2+m_4^2]} \ , 
\labbel{Sm} \end{equation}
with minor changes in the notation with respect to \cita{ER}.

The expansion in \( (n-4) \) of the solution of \Eq{ME} reads 
\begin{eqnarray} 
  \FF{4} &=& C^2(n) \Biggl\{ \frac{1}{(n-4)^2} \FFn{4}{-2} \nonumber \\ 
    && {\kern-140pt} + \frac{1}{(n-4)} \FFn{4}{-1} + \FFn{4}{0}
    + {\cal O} (n-4) \Biggr\} \ ,
\labbel{expn} \end{eqnarray} 
where the coefficient  
\begin{equation} 
C(n) = \left(2 \sqrt{\pi} \right)^{(4-n)} \Gamma\left(3-\frac{n}{2}\right) \ , 
\labbel{C} \end{equation} 

\noindent 
is not expanded to simplify analytical results. 
Similar Laurent-expansions in $(n-4)$ of $\F{j}$, (for $j=0,1,2,3$) and 
$\V$ were presented in \cita{CCLR1}, where 
explicit analytic expressions for the expansion coefficients 
$\Fn{j}{-2}$, $\Fn{j}{-1}$, $V^{(-2)}(m_1^2,m_2^2,m_3^2)$, 
$V^{(-1)}(m_1^2,m_2^2,m_3^2)$ and $V^{(0)}(m_1^2,m_2^2,m_3^2)$ are also given,
 while the $\Fn{j}{0}$ (for $j=0,1,2,3$) can be found numerically
 following the algorithm outlined in  \cita{CCR3}.
By substituting the expansions of all the amplitudes in the 
MDE \Eq{ME} and using the results of \cita{CCLR1}, 
the first coefficients are found to be 
\begin{eqnarray} 
    \FFn{4}{-2} &=& + \frac{1}{8} \nonumber \\ 
    \FFn{4}{-1} &=& - \frac{1}{16}  
                      - \frac{1}{2} S^{(0)}(m_1^2,m_4^2,p^2) \ , 
\labbel{9} \end{eqnarray} 
where \( S^{(0)}(m_1^2,m_4^2,p^2) \) is the finite part, at \( n=4 \), 
of the expansion of the MI from \Eq{Sm} 
\begin{equation} 
  \Sm = C(n) \Biggl\{ - \frac{1}{2} \frac{1}{(n-4)} 
       + S^{(0)}(m_1^2,m_4^2,p^2) + {\cal O} (n-4)\Biggr\} \ .
\labbel{12} \end{equation} 
Its analytic value is 
\begin{eqnarray} 
  S^{(0)}(m_1^2,m_4^2,p^2) &=& \frac{1}{2} - \frac{1}{4} \log(m_1 m_4) 
                                             \nonumber \\ 
   && {\kern-80pt} + \frac{1}{4p^2} \Biggl[ R(-p^2,m_1^2,m_4^2) 
   \log(u(p^2,m_1^2,m_4^2)) + (m_1^2-m_4^2)\log{\frac{m_1}{m_4}} \Biggl] 
\labbel{13} \ , \end{eqnarray} 
where \( R(-p^2,m_1^2,m_4^2) = \sqrt{ \RRp } \ \) and 
\[ u(p^2,m_1^2,m_4^2) = \frac 
     { \sqrt{ p^2+(m_1+m_4)^2 } - \sqrt{ p^2+(m_1-m_4)^2 } } 
     { \sqrt{ p^2+(m_1+m_4)^2 } + \sqrt{ p^2+(m_1-m_4)^2 } } \ . \] 
The function $\FFn{4}{0}$ is not known analytically, but it satisfies
the differential equation
\begin{eqnarray} 
  && {\kern-20pt} \RRp \ p^2 \ \frac{\partial}{\partial p^2} \FFn{4}{0} 
  = - (m_1^2+m_4^2) p^2 \FFn{4}{0} \nonumber \\ 
  && - (m_1^2-m_4^2)^2 \left( \FFn{4}{0} - \frac{1}{4} 
        S^{(0)}(m_1^2,m_4^2,p^2) \right)                  \nonumber \\ 
  && - (m_1^2-m_4^2) \Biggl[ 2 \Fn{0}{0} +m_1^2 \Fn{1}{0} \nonumber \\ 
  && +m_2^2 \Fn{2}{0} +m_3^2 \Fn{3}{0} 
       - V^{(0)}(m_2^2,m_3^2,m_4^2) \Biggr]               \nonumber \\ 
  && + \frac{m_1^2}{4} \left[ - \frac{3}{4} (m_2^2 + m_3^2) 
     + \frac{5}{4} m_4^2 - m_1^2 \right] + \frac{m_4^2}{16} 
       \left[ 3 (m_2^2 + m_3^2) - m_4^2 \right]           \nonumber \\ 
  && + \frac{(m_1^2 - m_4^2)}{8} \left[ \frac{3}{2} m_1^2 \log(m_1^2)
     + m_2^2 \log(m_2^2) + m_3^2 \log(m_3^2) 
     - \frac{1}{2} m_4^2 \log(m_4^2) \right]              \nonumber \\ 
  && + p^2 \Biggl[ 2 \Fn{0}{0} +3 m_1^2 \Fn{1}{0} 
     + m_2^2 \Fn{2}{0} \nonumber \\ 
  && +m_3^2 \Fn{3}{0} - V^{(0)}(m_2^2,m_3^2,m_4^2) 
     + \frac{15}{64} m_1^2 + \frac{3}{16} (m_2^2 + m_3^2) \nonumber \\ 
  && - \frac{3}{64} m_4^2 - \frac{3}{16} m_1^2 \log(m_1^2)
     - \frac{1}{8} \left( m_2^2 \log(m_2^2) + m_3^2 \log(m_3^2) \right)
     + \frac{1}{16} m_4^2 \log(m_4^2) \Biggr]             \nonumber \\ 
  && + p^4 \left( \frac{1}{64} -\frac{1}{4} 
        S^{(0)}(m_1^2,m_4^2,p^2) \right) \ .
\labbel{MEF40} \end{eqnarray} 
Although the coefficient $S^{(0)}(m_1^2,m_4^2,p^2)$ is given analytically
above, we report here also the differential equation, as it is used in the 
numerical program, for the related quantity 
$ \bar S^{(0)}(m_1^2,m_4^2,p^2) = S^{(0)}(m_1^2,m_4^2,p^2)
      -S^{(0)}(m_1^2,m_4^2,0)$, 
\begin{eqnarray} 
  && p^2 \frac{\partial}{\partial p^2} \bar S^{(0)}(m_1^2,m_4^2,p^2) = 
    -\bar S^{(0)}(m_1^2,m_4^2,p^2) \nonumber \\
  &&+\frac{p^2}{[p^2 + (m_1-m_4)^2]} \frac{1}{2} \Biggl\{
     \bar S^{(0)}(m_1^2,m_4^2,p^2) 
    +\frac{1}{4} \Biggl[ -1 + \frac{m_1 m_4}{(m_1^2 - m_4^2)} 
    \log\left( \frac{m_1^2}{m_4^2} \right) \Biggr] \Biggr\} \nonumber \\
  &&+\frac{p^2}{[p^2 + (m_1+m_4)^2]} \frac{1}{2} \Biggl\{ 
     \bar S^{(0)}(m_1^2,m_4^2,p^2) 
    +\frac{1}{4} \Biggl[ -1 - \frac{m_1 m_4}{(m_1^2 - m_4^2)} 
     \log\left( \frac{m_1^2}{m_4^2} \right) \Biggr] \Biggr\} \ .
\labbel{MES0} \end{eqnarray} 
%

\section{Initial conditions (expansion at \( p^2 \simeq 0 \) ).} 

To start the Runge-Kutta advancing solution method we choose 
as initial point the 
{\it special point} $p^2=0$, which allows the analytic calculation.
However, as this is one of the points where the coefficient multiplying  
the derivative of $F_4^{(0)}$ vanishes (\Eq{MEF40}), it is necessary  
to know also the second term in the expansion at \( p^2 \simeq 0 \). 

The general massive 1-loop self-mass expansion at \( p^2 \simeq 0 \), 
was presented in \cita{ER}, but we report here the explicit formulae to 
uniform the notation
\begin{equation} 
  S^{(0)}(m_1^2,m_4^2,p^2) = S_{0}^{(0)}(m_1^2,m_4^2) 
       + p^2 S^{(0)}_1(m_1^2,m_4^2) + {\cal O}((p^2)^2)  \ ,
\labbel{S0p0} \end{equation} 
where 
\begin{eqnarray} 
  S_{0}^{(0)}(m_1^2,m_4^2) \equiv
 S^{(0)}(m_1^2,m_4^2,0) &=& -\frac{1}{4}          \left[  
 \frac{m_1^2}{(m_1^2 - m_4^2)} \log(m_1^2)
-\frac{m_4^2}{(m_1^2 - m_4^2)} \log(m_4^2) - 1 \right] , \nonumber \\
  S^{(0)}_1(m_1^2,m_4^2) &=&  \frac{1}{4} \Biggl[ 
   \frac{m_1^2 m_4^2}{(m_1^2-m_4^2)^3} \log\left( \frac{m_1^2}{m_4^2} \right) 
  -\frac{m_1^2+m_4^2}{2(m_1^2-m_4^2)^2} \Biggr] \ . 
\labbel{S00} \end{eqnarray}

\noindent
As $p^2=0$ is a regular point, we define
the expansion at \( p^2 \simeq 0 \) of the MI from \Eq{MI} as 
\begin{eqnarray} 
  F_4(n,m_1^2,m_2^2,m_3^2,m_4^2,p^2) &=& 
     F_{4,0}(n,m_1^2,m_2^2,m_3^2,m_4^2) \nonumber \\
 &+& F_{4,1}(n,m_1^2,m_2^2,m_3^2,m_4^2) \ p^2 +{\cal O}((p^2)^2)
\labbel{F4p0} \end{eqnarray} 
The value at \( p^2 = 0 \) is easily found from \Eq{MI} and reads

\begin{eqnarray} 
 F_{4,0}(n,m_1^2,m_2^2,m_3^2,m_4^2) &\equiv& 
 F_4(n,m_1^2,m_2^2,m_3^2,m_4^2,0) \nonumber \\  
 &=& \frac{V(n,m_2^2,m_3^2,m_4^2)-V(n,m_1^2,m_2^2,m_3^2)}{m_1^2-m_4^2} \ .
\labbel{F40} \end{eqnarray} 
It can be in turn expanded in $(n-4)$ in the usual way
\begin{eqnarray} 
 && F_{4,0}(n,m_1^2,m_2^2,m_3^2,m_4^2) = C^2(n) \Biggl\{ 
     \frac{1}{(n-4)^2} F_{4,0}^{(-2)}(m_1^2,m_2^2,m_3^2,m_4^2) \nonumber \\
 &&{\kern-15pt} + \frac{1}{(n-4)}   F_{4,0}^{(-1)}(m_1^2,m_2^2,m_3^2,m_4^2) 
    + F_{4,0}^{(0)}(m_1^2,m_2^2,m_3^2,m_4^2) +{\cal O}(n-4) \Biggr\} \ , 
\labbel{F40n} \end{eqnarray} 
with
\begin{eqnarray} 
    F_{4,0}^{(-2)}(m_1^2,m_2^2,m_3^2,m_4^2) &=& + \frac{1}{8} \nonumber \\ 
    F_{4,0}^{(-1)}(m_1^2,m_2^2,m_3^2,m_4^2) &=& - \frac{1}{16}  
                      - \frac{1}{2} S_{0}^{(0)}(m_1^2,m_4^2)    \nonumber \\
    F_{4,0}^{(0)}(m_1^2,m_2^2,m_3^2,m_4^2)  &=& 
 \frac{V^{(0)}(m_2^2,m_3^2,m_4^2)-V^{(0)}(m_1^2,m_2^2,m_3^2)}{m_1^2-m_4^2} \ . 
\labbel{F40nc} \end{eqnarray} 
The expansion in $(n-4)$ of the coefficient of $p^2$ in \Eq{F4p0} is
\begin{eqnarray} 
 F_{4,1}(n,m_1^2,m_2^2,m_3^2,m_4^2) &=& C^2(n) \Biggl\{ 
     \frac{1}{(n-4)^2} F_{4,1}^{(-2)}(m_1^2,m_2^2,m_3^2,m_4^2) \nonumber \\
     + \frac{1}{(n-4)}   F_{4,1}^{(-1)}(m_1^2,m_2^2,m_3^2,m_4^2) 
     && + F_{4,1}^{(0)}(m_1^2,m_2^2,m_3^2,,m_4^2) +{\cal O}(n-4) \Biggr\} \ , 
\labbel{F41n} \end{eqnarray} 
with
\begin{eqnarray} 
  F_{4,1}^{(-2)}(m_1^2,m_2^2,m_3^2,m_4^2) &=& 0                  \nonumber \\ 
  F_{4,1}^{(-1)}(m_1^2,m_2^2,m_3^2,m_4^2) &=& - \frac{1}{2} 
  S^{(0)}_1(m_1^2,m_4^2)         \nonumber \\
  F_{4,1}^{(0)}(m_1^2,m_2^2,m_3^2,m_4^2)  &=& 
  \frac{m_1^2}{(m_1^2 -m_4^2)^2 \Rs} \Biggl\{                    \nonumber \\
  &&{\kern-120pt} \frac{m_2^2 (m_3^2 -m_1^2 -m_2^2)}{8} \Biggl[ 7 
  +\frac{1}{2} \log(m_1^2) \log\left( \frac{m_3^2}{m_2^2} \right)
  +\log(m_2^2) \left( \log(m_3^2) -\frac{3}{2} +\frac{1}{4} \log(m_2^2)\right)
  \nonumber \\
  &&+\log(m_3^2) \frac{3}{2} 
  \left( -3 +\frac{1}{2} \log(m_3^2) \right) \Biggr]             \nonumber \\
  &&{\kern-120pt} +\frac{m_2^4}{8} \Biggl[ 7 +\frac{1}{2} \log^2(m_2^2 m_3^2) 
  -3 \log(m_2^2 m_3^2) \Biggr]                                   \nonumber \\
  &&{\kern-120pt} +\frac{m_1^2 m_2^2}{4} \Biggl[ 7 +\log(m_1^2) \left( 
  -\frac{1}{2}\log(m_2^2) +\log(m_3^2) -\frac{3}{2} +\frac{1}{4}\log(m_1^2) 
  \right)                                                        \nonumber \\
  && +\log(m_3^2) \left( \frac{1}{2} \log(m_2^2) -\frac{9}{2} 
            +\frac{3}{4} \log(m_3^2) \right) \Biggr]             \nonumber \\
  &&{\kern-120pt} +\frac{m_1^2 (m_3^2 -m_1^2 -m_2^2)}{16} 
  \Biggl[ 7 +\frac{1}{2} \log^2(m_1^2 m_3^2) -3 \log(m_1^2 m_3^2 ) \Biggr] 
                                                                 \nonumber \\
  &&{\kern-120pt} + (m_3^2 -m_1^2 +m_2^2) \ V^{(0)}(m_1^2,m_2^2,m_3^2) 
  \ \Biggr\}                                                     \nonumber \\
  &&{\kern-120pt} - \frac{1}{(m_1^2-m_4^2)} \Bigg\{ \frac{1}{32}
  \left( -\frac{3}{4} + \log(m_4^2) \right) 
  + F_{0,1}^{(0)}(m_1^2,m_2^2,m_3^2)    \nonumber \\
  &&{\kern-120pt} + \frac{m_1^2}{2} F_{1,1}^{(0)}(m_1^2,m_2^2,m_3^2)
                  + \frac{m_2^2}{2} F_{2,1}^{(0)}(m_1^2,m_2^2,m_3^2) 
                  + \frac{m_3^2}{2} F_{3,1}^{(0)}(m_1^2,m_2^2,m_3^2) 
  \Bigg\}                                                        \nonumber \\
  &&{\kern-120pt} +\frac{m_1^2}{(m_1^2-m_4^2)^2} \frac{1}{4}
  \Biggl\{ 1 -\frac{3}{8} \log(m_1^2) -\frac{3}{4} \log(m_3^2) + \frac{1}{8} 
  \left[ \log(m_4^2) +\log^2(m_1^2 m_3^2) \right] \Biggr\}       \nonumber \\
   &&{\kern-120pt} + \frac{m_1^2}{(m_1^2-m_4^2)^3} \left[ 
   V^{(0)}(m_1^2,m_2^2,m_3^2) - V^{(0)}(m_2^2,m_3^2,m_4^2) \right] \ .
\labbel{F41nc} \end{eqnarray} 
The above expression for $F_{4,1}^{(0)}(m_1^2,m_2^2,m_3^2,m_4^2)$, 
although not at a glance, is symmetric in the exchange of $m_2^2$ and $m_3^2$.

The coefficient of the double expansion, at $n \simeq 4$ and 
$ p^2 \simeq 0 $, $F_{0,1}^{(0)}(m_1^2,m_2^2,m_3^2)$ is explicitly given 
in Eq.(51) of \cita{CCLR1}, while the other coefficients 
$F_{1,1}^{(0)}(m_1^2,m_2^2,m_3^2)$, 
$F_{2,1}^{(0)}(m_1^2,m_2^2,m_3^2)$, 
$F_{3,1}^{(0)}(m_1^2,m_2^2,m_3^2)$, 
can be easily found by using the relations 
$F_i = -\frac{\partial}{\partial m_i^2} F_0, ( i =1,2,3)$.

\section{Numerics.} 

As already illustrated for the case of the 2-loops sunrise graph in 
\cita{CCR3}, one can use the Runge-Kutta method \cita{PTVF} to advance the 
solution of the MDE, in this particular case \Eq{MEF40},
 from the known initial conditions at $p^2 \simeq 0$, 
following a complex $p^2$-path in the lower half-plane. 
We add \Eq{MEF40} and \Eq{MES0} to the system of the four MDE for the four 
sunrise MI's~\cita{CCR3}, as they are present also in \Eq{MEF40} , and 
we solve the system at the same time for all the MI's in the same 
numerical program. 

All the features discussed in \cita{CCR3} regarding precision and organization 
of the program remain unchanged. 
Of course the execution time increases as the system of equations is bigger. 

Again for convenience, we use reduced masses and a reduced external momentum
 squared 
\begin{equation} 
m_{i,r} \equiv \frac{m_i}{\mu}, \quad, \quad 
p_r^2 \equiv \frac{p^2}{\mu^2}, \quad, \quad  \mu=m_1+m_2+m_3 \ , 
\labbel{red1} \end{equation} 
where the choice for $\mu$ is motivated by faster numerical calculations.

As discussed in \cita{CCR3}, close to a {\it special point}, 
different from $p_r^2 = 0$, numerical problems 
arise.
For $F^{(0)}_{4}(m_1^2,m_2^2,m_3^2,m_4^2,p_r^2)$ 
the {\it special points} are: $p_r^2 = \infty$ 
(analytically discussed in \cita{CCLR2}), $p_r^2 = 0$, the 
2-body threshold  $p_r^2 \equiv p_{2b,th,r}^2 = -(m_{1,r}+m_{4,r})^2$, 
the 3-body threshold  
$p_r^2 \equiv p_{3b,th,r}^2 = -(m_{1,r}+m_{2,r}+m_{3,r})^2$ 
and the 2-body and 3-body pseudo-thresholds  
$p_r^2 \equiv p_{2b,ps,r}^2 = -(m_{1,r}-m_{4,r})^2$, 
$p_r^2 \equiv p_{3b,ps1,r}^2 = -(m_{1,r}+m_{2,r}-m_{3,r})^2$, 
$p_r^2 \equiv p_{3b,ps2,r}^2 = -(m_{1,r}-m_{2,r}+m_{3,r})^2$, 
$p_r^2 \equiv p_{3b,ps3,r}^2 = -(m_{1,r}-m_{2,r}-m_{3,r})^2$. 
To obtain values of the MI's for $p_r^2$
 close to thresholds and pseudo-thresholds we 
use the method discussed in detail in the next section.

\begin{figure}[tbp]
\begin{center}
\epsfig{file=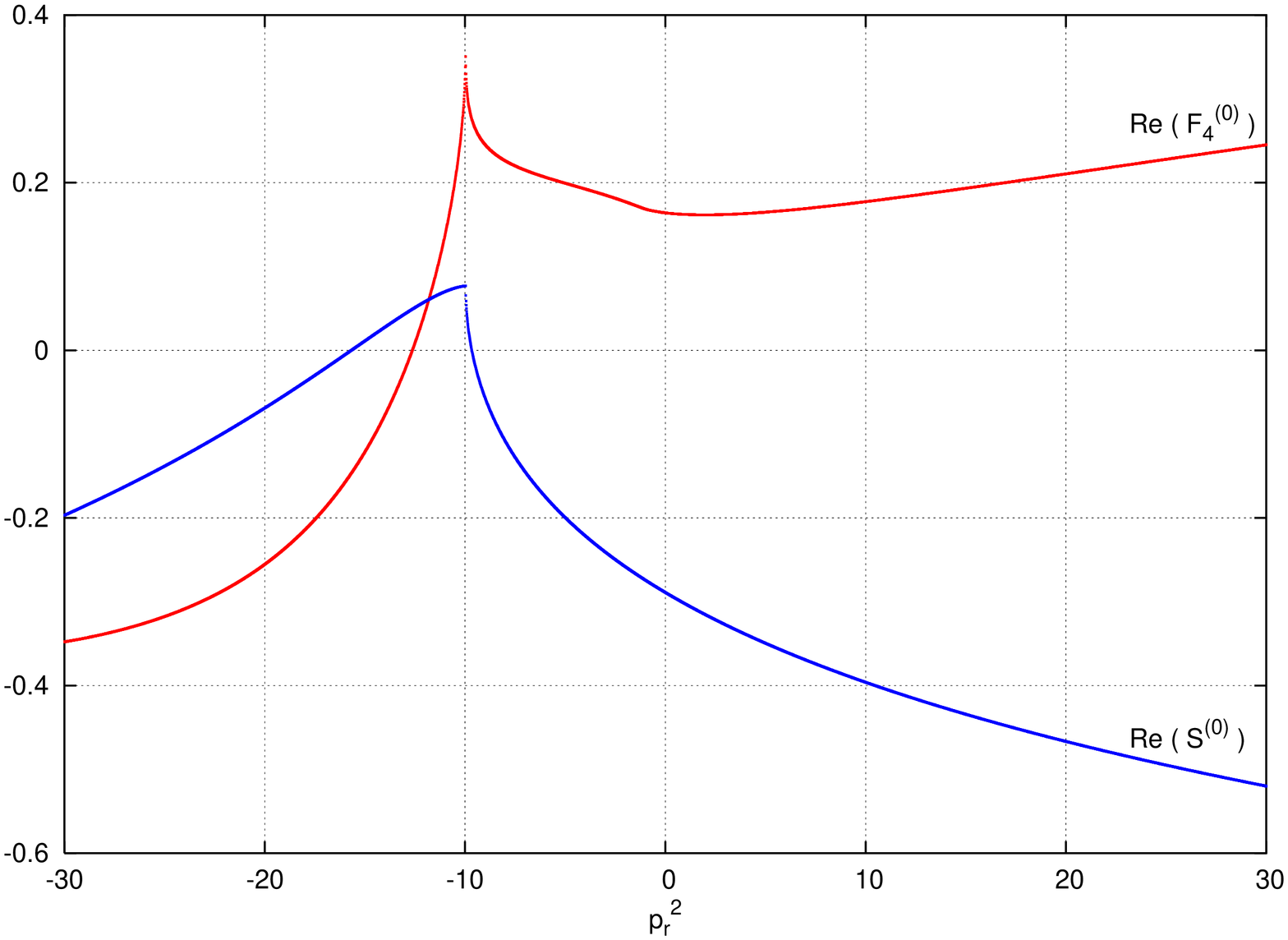,height=\textwidth,width=7cm,angle=-90}
\end{center}
\caption{ Plots of  
$\Re F^{(0)}_{4}(m_1^2,m_2^2,m_3^2,m_4^2,p_r^2)$ and 
$\Re S^{(0)}(m_1^2,m_4^2,p_r^2)$ 
as a function of \(p_r^2\) 
for $m_1=2$, $m_2=1$, $m_3=4$, $m_4=20.1$ and $\mu = m_1+m_2+m_3$. }
\label{fig:f1}
\end{figure}

\begin{figure}[tbp]
\begin{center}
\epsfig{file=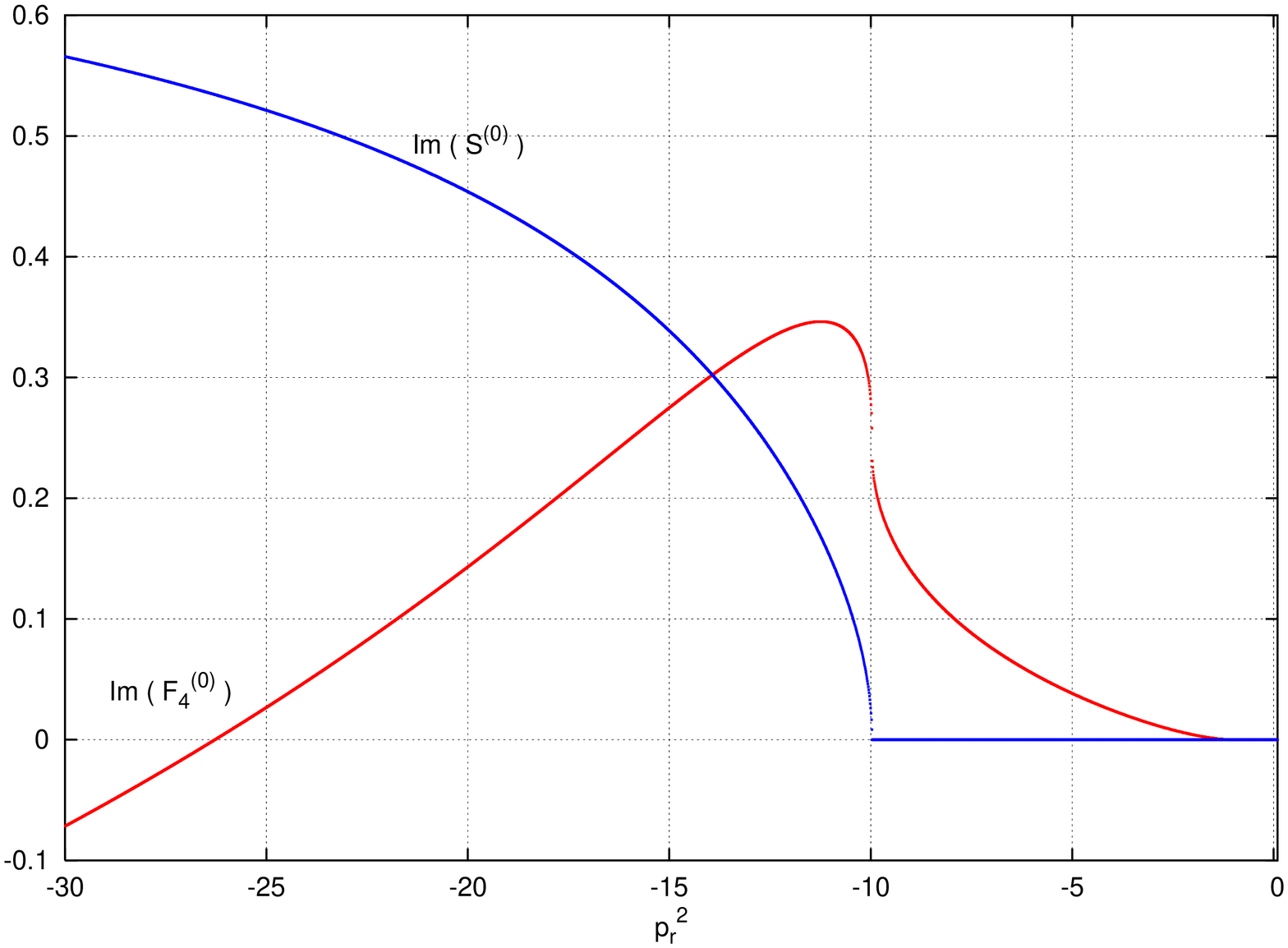,height=\textwidth,width=7cm,angle=-90}
\end{center}
\caption{ Plots of  
$\Im F^{(0)}_{4}(m_1^2,m_2^2,m_3^2,m_4^2,p_r^2)$ and 
$\Im S^{(0)}(m_1^2,m_4^2,p_r^2)$ 
as a function of \(p_r^2\) 
for $m_1=2$, $m_2=1$, $m_3=4$, $m_4=20.1$ and $\mu = m_1+m_2+m_3$. }
\label{fig:f2}
\end{figure}

To illustrate some results of the program, and in the same time to show
 the behaviour of the function $F_4$ over a wide range of the variable
 $p_r^2$, we present here plots for two different sets of masses.
In Fig. {\ref{fig:f1}} we plot the values of 
$\Re F^{(0)}_{4}$ and $\Re S^{(0)}$ 
as functions of $p_r^2$, for the choice of the mass-values 
$m_1=2$, $m_2=1$, $m_3=4$, $m_4=20.1$ (arbitrary units).
The 2-body threshold, $p^2=-(m_1+m_4)^2)\ ,$ corresponds to 
$p_{2b,th,r}^2=-9.96755102$. The presence of the 3-body threshold
($p_{3b,th,r}^2=-1,\ $ due to the very definition of the scale $\mu$)
is hardly visible from the $\Re F^{(0)}_{4}$ plot, but clearly from its
 imaginary part plotted in Fig.{\ref{fig:f2}}. Its behaviour 
 around the 2-body threshold is more `fuzzy' due to the presence of root terms
 in the threshold expansion
 (see next section for an explicit form of the threshold expansion).
The strikingly different behaviour of $F_4^{(0)}$ and $S^{(0)}$ near the
 2-body threshold is due to additional logarithmic terms in the expansion
 of $F_4^{(0)}$, which are absent in $S^{(0)}$ expansion.

\begin{figure}[tbp]
\begin{center}
\epsfig{file=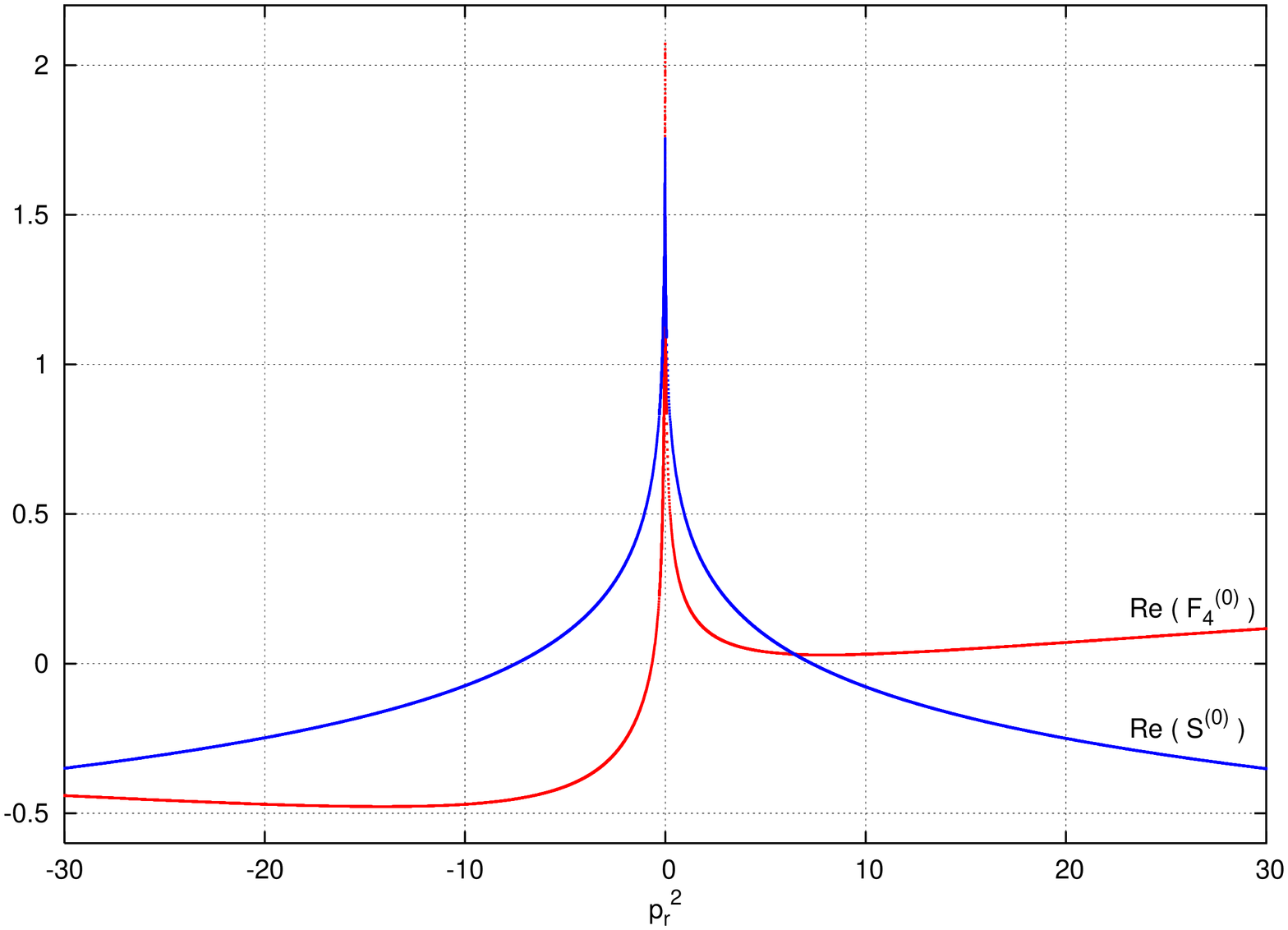,height=\textwidth,width=7cm,angle=-90}
\end{center}
\caption{ Plots of  
$\Re F^{(0)}_{4}(m_1^2,m_2^2,m_3^2,m_4^2,p_r^2)$ and 
$\Re S^{(0)}(m_1^2,m_4^2,p_r^2)$ 
as a function of \(p_r^2\) 
for $m_1=1$, $m_2=9$, $m_3=200$, $m_4=20.1$ and $\mu = m_1+m_2+m_3$. }
\label{fig:f3}
\end{figure}

In Fig. {\ref{fig:f3}} are plotted the values of $\Re F^{(0)}_{4}$ and 
$\Re S^{(0)}$ as a function of $p_r^2$, for the second choice of the 
mass-values, where one of the masses is considerably bigger than the others 
$m_1=1$, $m_2=9$, $m_3=200$, $m_4=20.1$.  This time the higher threshold 
is the 2-body threshold at $ p_{2b,th,r}^2 = -0.010095465$, while the
3-body threshold is as always at $p_{3b,th,r}^2 = -1$. 
Both functions show pronounced peaks around the 2-body threshold, 
and the 'zoom' of the peaks in Fig. {\ref{fig:f4}}
shows that the qualitative behaviour of both functions nearby the 2-body
 threshold does not depend on masses (compare Fig. {\ref{fig:f1}}).

\begin{figure}[tbp]
\begin{center}
\epsfig{file=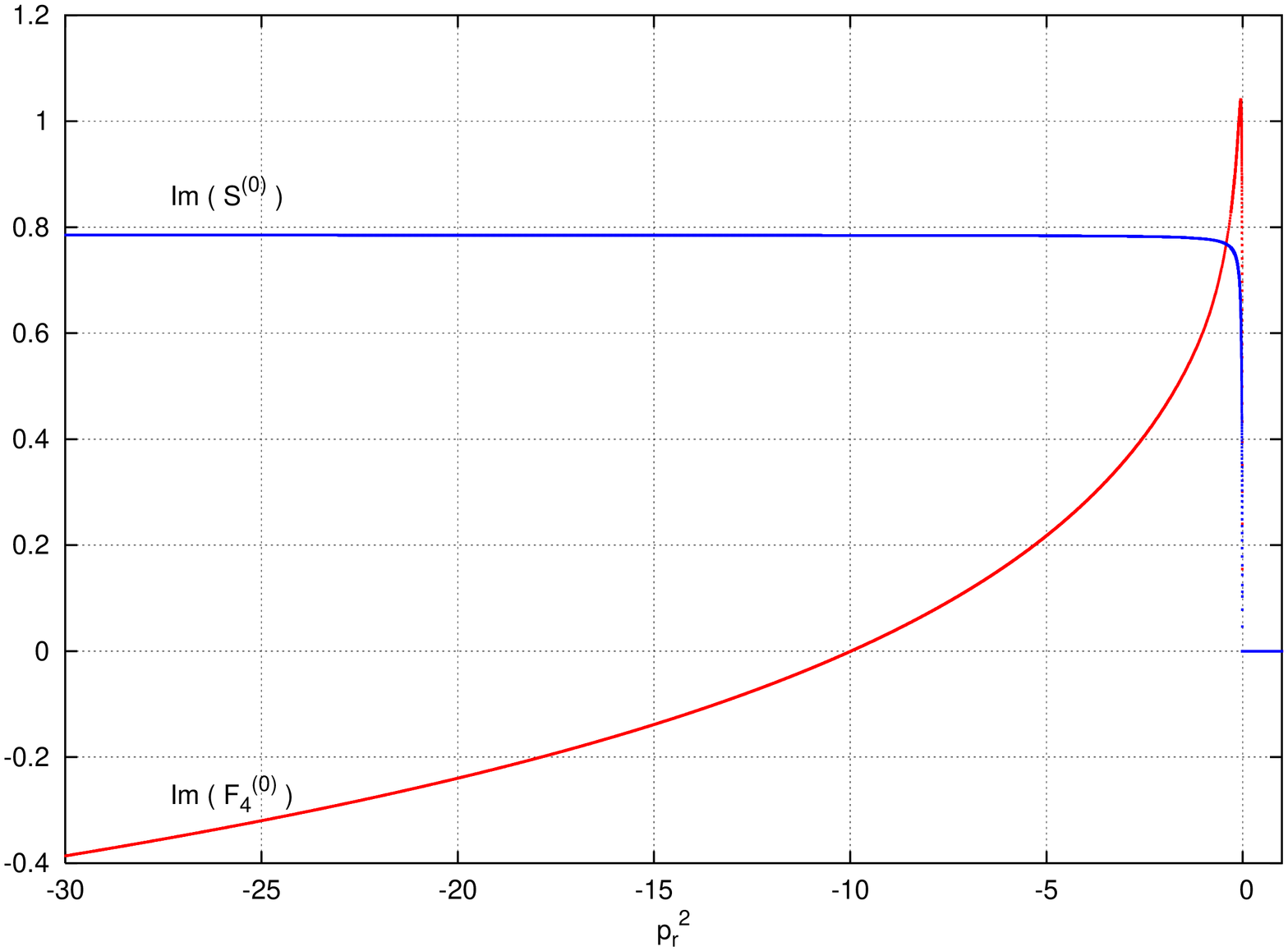,height=\textwidth,width=7cm,angle=-90}
\end{center}
\caption{ Plots of  
$\Im F^{(0)}_{4}(m_1^2,m_2^2,m_3^2,m_4^2,p_r^2)$ and 
$\Im S^{(0)}(m_1^2,m_4^2,p_r^2)$ 
as a function of \(p_r^2\) 
for $m_1=1$, $m_2=9$, $m_3=200$, $m_4=20.1$ and $\mu = m_1+m_2+m_3$. }
\label{fig:f5}
\end{figure}

\begin{figure}[tbp]
\begin{center}
\epsfig{file=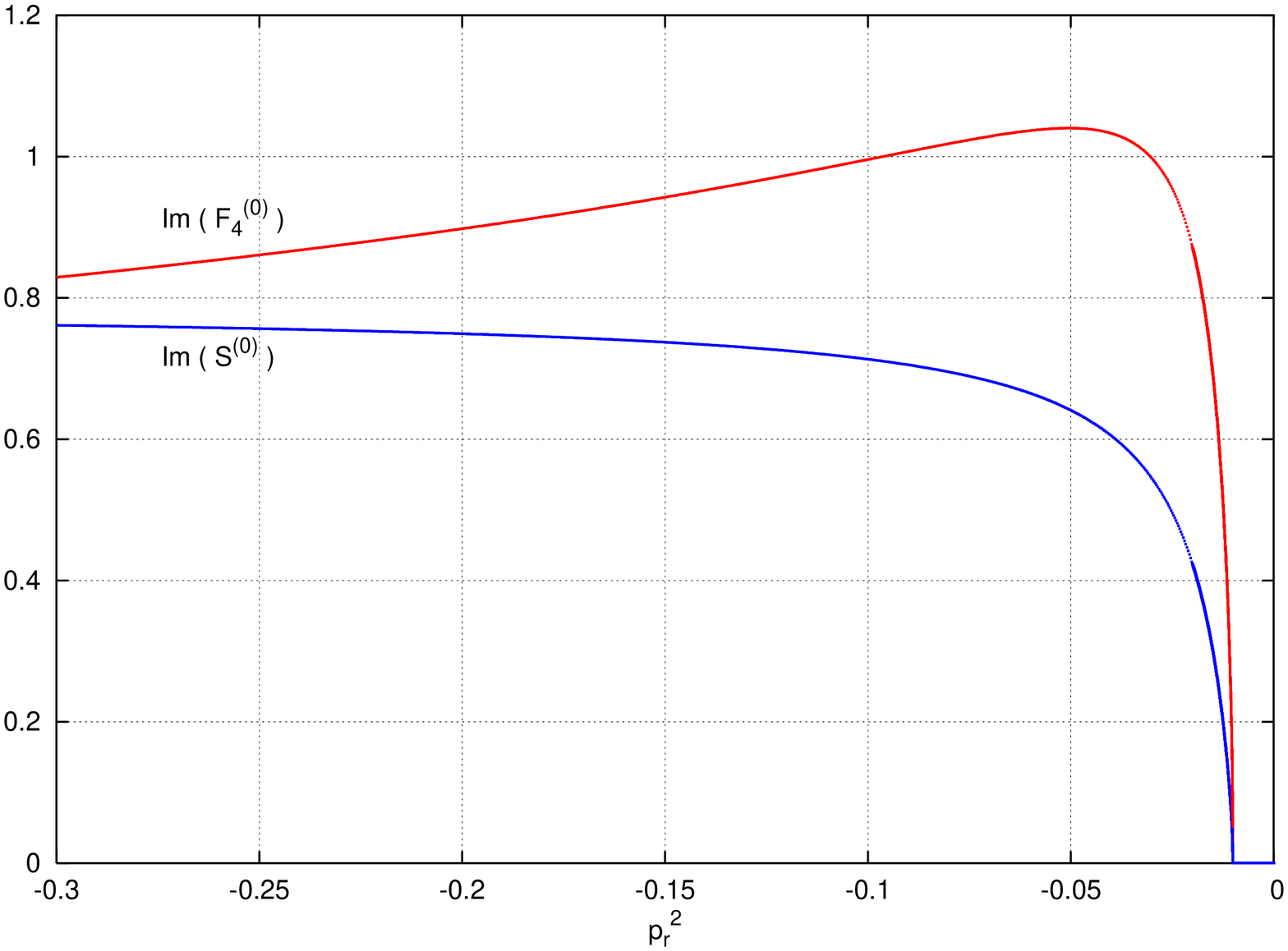,height=\textwidth,width=7cm,angle=-90}
\end{center}
\caption{ Enlargements for the peaks of the plots of  
$\Im F^{(0)}_{4}(m_1^2,m_2^2,m_3^2,m_4^2,p_r^2)$ and 
$\Im S^{(0)}(m_1^2,m_4^2,p_r^2)$ 
as a function of \(p_r^2\) 
for $m_1=1$, $m_2=9$, $m_3=200$, $m_4=20.1$ and $\mu = m_1+m_2+m_3$. }
\label{fig:f6}
\end{figure}

The imaginary part of $\Im F^{(0)}_{4}$ starts 
to be non vanishing at the 2-body threshold, as shown in Fig. {\ref{fig:f5}}, 
but this time no visible modification comes in at the 3-body threshold.
The peaking behaviour of the imaginary parts is shown enlarged in 
Fig. {\ref{fig:f6}}.

In Table {\ref{tab:tab1}} are reported the benchmark values of 
$\Re F^{(0)}_{4}(m_1^2,m_2^2,m_3^2,m_4^2,p_r^2)$ and 
$\Im F^{(0)}_{4}(m_1^2,m_2^2,m_3^2,m_4^2,p_r^2)$ for 
the masses \(m_1 = 1 \), \(m_2 = 9 \), \(m_3 = 200 \), 
\(m_4 = 20.1 \), \(\mu = m_1+m_2+m_3 \). 

The values at thresholds and pseudo-thresholds are obtained with the 
method discussed in the next section, the value at $p_r^2=0$ is from 
the analytic formula.
\begin{table}
\begin{center}
\begin{tabular}{llll}
 & $p^2_r$ & $\Re F^{(0)}_{4}$ & $\Im F^{(0)}_{4}$ \\ \hline  
 &{\small -30.}                      &          {\small -0.44092793813(5)}
                                     &          {\small -0.38633149701(4)}\\
 &{\small -15.}                      &          {\small -0.47753230147(5)}
                                     &          {\small -0.13848347339(5)}\\
 &{\small -1.5}                      &          {\small -0.19215624308(2)}
                                     & \kern+3pt{\small  0.5235669521(1) }\\
{\small\(p_{3b,th,r}^2\) =}
 &{\small -1.0}                      &          {\small -0.0990578391(1) }
                                     & \kern+3pt{\small  0.60473101562(7)}\\
 &{\small -0.99}                     &          {\small -0.09669632676(2)}
                                     & \kern+3pt{\small  0.6066658942(1) }\\
{\small\(p_{3b,ps3,r}^2\)=}
 &{\small -0.981043}                 &          {\small -0.094558472357(8)}
                                     & \kern+3pt{\small  0.60841511708(6)}\\
 &{\small -0.9}                      &          {\small -0.07412586818(2)}
                                     & \kern+3pt{\small  0.6249850792(1) }\\ 
{\small\(p_{3b,ps2,r}^2\)=}
 &{\small -0.835918}                 &          {\small -0.056354400576(6)}
                                     & \kern+3pt{\small  0.63914196276(5)}\\
 &{\small -0.825}                    &          {\small -0.05316178728(1)}
                                     & \kern+3pt{\small  0.6416578704(1) }\\
{\small\(p_{3b,ps1,r}^2\)=}
 &{\small -0.818594}                 &          {\small -0.051264529466(5)}
                                     & \kern+3pt{\small  0.64314892655(5)}\\
 &{\small -0.8}                      &          {\small -0.04565273027(1)}
                                     & \kern+3pt{\small  0.6475413524(1) }\\
 &{\small -0.1}                      & \kern+3pt{\small  0.65325134562(1)}
                                     & \kern+3pt{\small  0.99590695379(1)}\\
{\small\(p_{2b,th,r}^2\) =}
 &{\small -0.010095}                 & \kern+3pt{\small  2.101104208(4)}
                                     & \kern+3pt{\small  0.0             }\\
{\small\(p_{2b,ps,r}^2\) =}
 &{\small -0.008272}                 & \kern+3pt{\small  1.697672304903(6)}
                                     & \kern+3pt{\small  0.0             }\\
 &{\small 0.0}                         & \kern+3pt{\small  1.351553692518424}
                                     & \kern+3pt{\small  0.0             }\\
 &{\small 1.0}                         & \kern+3pt{\small  0.21346230166(6)}
                                     & \kern+3pt{\small  0.0             }\\
 &{\small 30.0}                        & \kern+3pt{\small  0.117099162908(6)}
                                     & \kern+3pt{\small  0.0             }\\
\hline
\end{tabular}
\end{center}
\caption{ The benchmark values of 
$\Re F^{(0)}_{4}(m_1^2,m_2^2,m_3^2,m_4^2,p_r^2)$ and 
$\Im F^{(0)}_{4}(m_1^2,m_2^2,m_3^2,m_4^2,p_r^2)$ for 
the masses \(m_1 = 1 \), \(m_2 = 9 \), \(m_3 = 200 \), 
\(m_4 = 20.1 \), \(\mu = m_1+m_2+m_3 \). }
\label{tab:tab1}
\end{table}

\section{Thresholds and pseudo-thresholds.}

The numerical calculation with the system of MDE does not allow for 
$p_r^2$ much closer than $10^{-4}$ to the thresholds and 
pseudo-thresholds ({\it special points}), due to the numerical 
instability of the equations in these {\it surrounding-domains}. 
So a special treatment is required to get precise values of the MI's there. 

For the sunrise MI's the analytical expansions at pseudo-thresholds 
\cite{CCR1} and at threshold \cite{CCR2} were previously obtained.
In \cita{CCR3} these results were used as starting points for the system 
of MDE's for the sunrise MI's to get reliable results in the 
{\it surrounding-domain} of these points.
However this method is not universal, as it requires a separate analytic
calculation of the MI expansion at these points, which is difficult and 
probably not always possible.

We discuss here in detail the features, advantages and limitations of an 
approximation method, which is rather precise, universal and easy to use. 
This method was recently proposed in \cite{Martin}. However the 'universal'
 approximant suggested by the author applies only to some cases and 
fails at the 2-body threshold relevant for the present calculations.

The method consists in the construction of a suitable approximant of the MI, 
which due to the smallness of the {\it surrounding-domain} is naturally the 
expansion of the MI around the considered {\it special point}. 
The proper form of the expansion (with undetermined coefficients) around each 
{\it special point} can easily be deduced from the MDE's 
themselves~\cite{Ince}.

The coefficients of the approximant are calculated using numerical values 
of the MI, obtained solving the system of MDE for some points nearby the 
{\it special point}, 
but outside its {\it surrounding-domain} to avoid numerical instability. 
The approximant is used to get the values of the MI at the 
{\it special point} and within its {\it surrounding-domain}.
Note that the proper form of the expansion around a {\it special point} is 
crucial to produce right results around {\it special points}.
With the wrong choice of the approximant one may still get the values
at the {\it special points} right, when the points chosen for the 
 approximant `construction' are symmetrically distributed
 around the {\it special point}. The values obtained 
 for  the {\it surrounding-domains}
 will be however wrong.

To test the precision of this approximation method, we compare 
its result at the {\it special point} with that obtained from analytic result 
there, when available, and in the {\it surrounding-domain} with 
the result given by the advanced numerical solution of the MDE's started 
from the {\it special point}.
Of course the approximation method can be used also in a region where a MI 
does not contain a {\it special point} to test the precision of the method. 

The 2 and 3-body pseudo-thresholds are the simplest points to be treated 
in this way: as they are regular points the approximant can simply be 
the power expansion.
For each MI (\(F_i^{(0)}, \ i=0,1,2,3,4\) and \(S^{(0)}\)) in each of 
its pseudo-thresholds can be used an approximant of the type
\begin{eqnarray}
F(x) = a_0 +a_1 x +a_2 x^2 +a_3 x^3 \ , 
\labbel{app1} 
\end{eqnarray}
where \(x = p_r^2 - p_{ps,r}^2\) and  \(p_{ps,r}^2\) is the value of $p_r^2$ 
at the pseudo-threshold. 
As we want to use \(x\) well below \(10^{-4}\), in \Eq{app1} it is enough 
to use four terms in the expansion, the truncation causing a relative error 
of the order of \(x^4 = 10^{-16}\). 
The actual implementation of \(F(x)\) uses the four points  
\(x = \pm 10^{-4},\pm 0.5 \cdot 10^{-4}\) to calculate the four coefficients 
\(a_i, \ \ i=0,1,2,3 \), the error associated to its value is dominated
by the error coming from the numerical calculation of the MI at these points. 
It has to be noted that the coefficients of the expansion \Eq{app1} can be 
complex if the expanded function is complex valued -- as it is the case 
above threshold(s). 

The values of the sunrise MI's \( \ F_i^{(0)}, \ i=0,1,2,3 \) at the 3-body 
pseudo-thresholds and of the 1-loop self-mass MI \(\ S^{(0)}\) at the 2-body 
pseudo-threshold, can be obtained from known analytical formulas 
\cite{CCR1,ER}. 
The approximation method can reach in the implemented program a relative 
error of the order of \( 10^{-11}-10^{-12}\) and its results are always 
in agreement with the analytical ones within this error. 

In the {\it surrounding-domain} (where \(x < 0.5\cdot10^{-4}\)) of the 
3-body pseudo-threshold approximation method results for the sunrise MI's 
are in agreement with the results obtained from MDE advancing solution 
starting from the pseudo-threshold \cita{CCR3} within a relative accuracy 
of the order of \( 10^{-9}\), which is the best accuracy obtainable by 
the MDE advancing solution program in this region.
However, as the approximation method values for \(F_i^{(0)}, \ i=0,1,2,3 \), 
and \(S^{(0)}\) reproduce well the values from analytical formulas at the 
pseudo-thresholds, and from MDE advancing solution program starting from 
$p_r^2=0$ in all tested points outside the {\it surrounding-domain} 
(where \(x \ge 0.5\cdot10^{-4}\)), we are confident that the approximation 
method is better and that the stated relative accuracy of order 
\( 10^{-11}-10^{-12}\) is a safe estimate.

At the 3-body threshold the proper approximants for the 
\(F_i^{(0)}, \ i=0,1,2,3,4\ \) are
\begin{eqnarray}
F_i^{(0)}(x) = a_{i,0} + \left(a_{i,1} +b_{i,1} \log(x)\right) x 
                       + \left(a_{i,2} +b_{i,2} \log(x)\right) x^2 
                       + \left(a_{i,3} +b_{i,3} \log(x)\right) x^3 \ , 
\labbel{app2} 
\end{eqnarray}
\noindent
where \(x = p_r^2 - p_{3b,th,r}^2\) 
and  \(p_{3b,th,r}^2 = -(m_{1,r}+m_{2,r}+m_{3,r})^2 = -1\).
We use 
\(x = \pm 0.5 \cdot 10^{-4}, \pm 10^{-4}, \pm 1.5 \cdot 10^{-4}, 
\pm 2 \cdot 10^{-4}\) to obtain numerically the coefficients 
\(a_{i,j}, \ i=0,1,2,3,4, \ j =0,1,2,3\) and \(b_{i,k}, \ k=1,2,3\). 
Again all the coefficients in the expansion of $F_4^{(0)}$ can be complex 
depending on the values of the 2-body and 3-body thresholds. For $S^{(0)}$ 
we use \Eq{app1} as it is regular at the 3-body threshold.

Comparisons with the analytical results for the sunrise MI at the 3-body 
threshold \cita{CCR3} and other tests like those at pseudo-thresholds 
were performed with the same conclusions, so again a precision of the 
order of \( 10^{-11}-10^{-12}\) can be assumed.

\begin{figure}[tbp]
\begin{center}
\epsfig{file=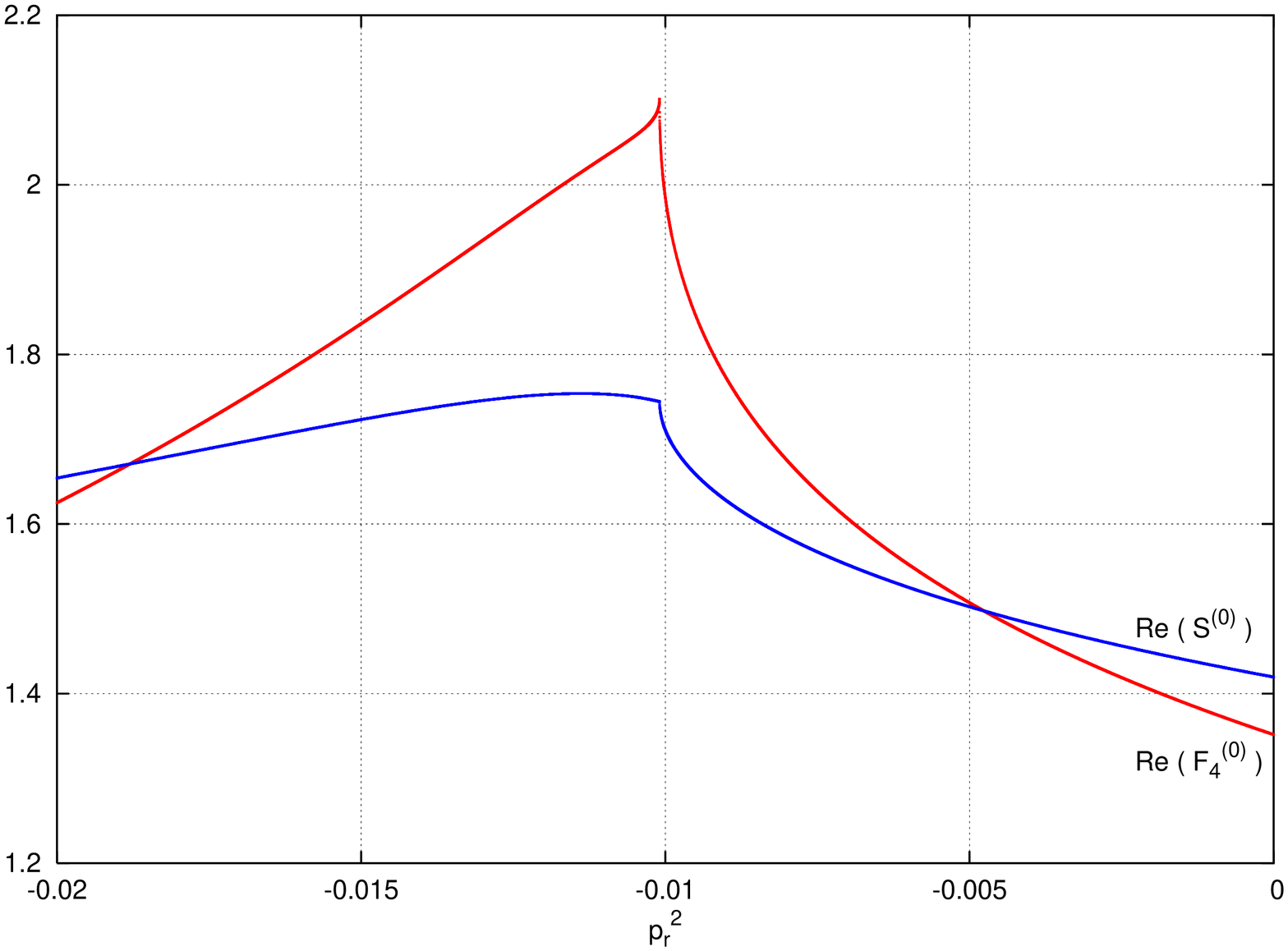,height=\textwidth,width=7cm,angle=-90}
\end{center}
\caption{ Enlargements for the peaks of the plots of  
$\Re F^{(0)}_{4}(m_1^2,m_2^2,m_3^2,m_4^2,p_r^2)$ and 
$\Re S^{(0)}(m_1^2,m_4^2,p_r^2)$ 
as a function of \(p_r^2\) 
for $m_1=1$, $m_2=9$, $m_3=200$, $m_4=20.1$ and $\mu = m_1+m_2+m_3$. }
\label{fig:f4}
\end{figure}

At the 2-body threshold the proper approximants for \(S^{(0)}\) and 
\(F_4^{(0)}\) are
\begin{eqnarray}
&&\kern-10ptS^{(0)} = a_{S,0} + a_{S,1} x + a_{S,2} x^2 + a_{S,3} x^3 
+\sqrt{x} \ \left(b_{S,0} + b_{S,1} x  + b_{S,2} x^2  + b_{S,3} x^3 \right) \ ,
\nonumber \\  
&&\kern-10ptF_4^{(0)} = a_{4,0} + a_{4,1} x + a_{4,2} x^2 + a_{4,3} x^3 
+\sqrt{x} \ \biggl[ b_{4,0} + b_{4,1}x + b_{4,2}x^2 + b_{4,3}x^3 \biggr] 
\nonumber \\
 &&\kern+14pt + \sqrt{x}\log(x) 
     \biggl[ c_{4,0} + c_{4,1}x + c_{4,2}x^2 + c_{4,3}x^3 \biggr] 
\labbel{app3} 
\end{eqnarray}
\noindent
where \(x = p_r^2 - p_{2b,th,r}^2\) 
and \(p_{2b,th,r}^2 = -(m_{1,r}+m_{4,r})^2\).
The threshold expansion can be easily derived from \Eq{MEF40} for 
\(F_4^{(0)}\), while for \(S^{(0)}\) it was given in \cita{ER}. 
The real constants \(a_{S,j},\ b_{S,j}\) and real or complex constants 
\(a_{4,j},\ b_{4,j},c_{4,j},\ j=0,1,2,3\) (see discussion above) are obtained
calculating numerically with the MDE advancing solution 
the values of \(S^{(0)}\) at 
\(x=\pm i\cdot10^{-3}\), \(x=\pm i\cdot 0.5\cdot10^{-3}\), 
\(x=\pm i\cdot 0.25\cdot10^{-3}\), \(x=\pm i\cdot 0.125\cdot10^{-3}\),
and of \(F_4^{(0)}\) there and also at
\(x=\pm i\cdot 0.0625\cdot10^{-3}\), \(x=\pm i\cdot 0.03125\cdot10^{-3}\).
The choice of imaginary values for \(x\) allows to overcome the numerical 
problems arising in the present case for real values of \(x\) when 
\(|x|<10^{-4}\) with the usual MDE advancing solution program starting 
from $p_r^2=0$, due to the steep behaviour of the functions \(S^{(0)}\) 
and \(F_4^{(0)}\) nearby the 2-body threshold, as can be seen in 
Fig. {\ref{fig:f4}}.
On the other hand the MI values at \(|x| <10^{-4}\) are needed to get 
enough precise values for the coefficients in the approximants.
For \(F_i^{(0)},\  (i=0,1,2,3)\) we use \Eq{app1} as approximants.

For the \(S^{(0)}\) approximant values a relative precision of the order 
of \( 10^{-11}-10^{-12}\) is obtained as before, confirmed by the comparisons 
with the analytic result.

For the $F_4^{(0)}$ approximant values the contrasting requirements does 
not allow a relative precision better than \( 10^{-8}-10^{-9}\). 

\section{Comparisons.} 

Numerical values for the MI of \Eq{MI} for non-vanishing values of the
masses are reported in literature, only for small values of time-like $p^2$, 
in \cita{BT} and more extensively in \cita{BBBB}. 
More precisely there was calculated the quantity
\begin{eqnarray} 
 && T_{1234N}(\bar p^2;\bar m_1^2,\bar m_2^2,\bar m_3^2,\bar m_4^2) \equiv
  T_{1234}(\bar p^2;\bar m_1^2,\bar m_2^2,\bar m_3^2,\bar m_4^2,\mu^2) - 
  T_{1234}(\bar p^2;\bar m_1^2,\bar m_2^2,0,0,\mu^2)  \nonumber \\
 && = 16 F_4^{(0)}(m_1^2,m_2^2,m_3^2,m_4^2,p^2,\mu^2) - 
  T_{1234}^{(0)}(-p^2;m_4^2,m_1^2,0,0,\mu^2) 
   +\frac{1}{2} \zeta(2) \ , 
\labbel{comp} \end{eqnarray}
with $T-$functions defined in \cita{BT,BBBB}. 
The function 
$T_{1234N}(\bar p^2;\bar m_1^2,\bar m_2^2,\bar m_3^2,\bar m_4^2)$
is finite in the $n \to 4$ limit and it is independent from $\mu^2$, the 
arbitrary mass scale introduced in \Eq{MI}. 
Here the $\mu^2$ dependence is explicitly shown, whenever present. 
The quantity 
$T_{1234}(\bar p^2;\bar m_1^2,\bar m_2^2,\bar m_3^2,\bar m_4^2,\mu^2)$
is like the MI of \Eq{MI}, with a different normalization factor, which 
results in the factor $16$ and the additive 
quantity $\zeta(2)/2$,  the second line of \Eq{comp}.
The analytical expression for 
$T_{1234}(\bar p^2;\bar m_1^2,\bar m_2^2,0,0,\mu^2)$
is given in \cita{ST}.

Due to the different assignment of the mass indexes ($m_i$) to the lines 
in our Fig.{\ref{fig:f0}} with respect to the correspondent ones ($\bar m_j$) 
in \cita{BT,BBBB,ST}, we have the correspondence $m_1=\bar m_2=3$, 
$m_2=\bar m_4=7$, $m_3=\bar m_3=5$, $m_4=\bar m_1=1$, where the numerical 
values are for the comparison with Table 1 of \cita{BBBB}. 
One has also $\bar p^2=-p^2$, due to Minkowski-Euclidean space transformation. 
In the second line of \Eq{comp} 
$F_4^{(0)}(m_1^2,m_2^2,m_3^2,m_4^2,p^2,\mu^2)$
is calculated through the differential equation \Eq{MEF40} and 
$T_{1234}^{(0)}(-p^2;m_4^2,m_1^2,0,0,\mu^2)$ is the finite part
in the $n \to 4$ limit of 
$T_{1234}(-p^2;m_4^2,m_1^2,0,0,\mu^2)$ from \cita{ST}. 

\begin{table}
\begin{center}
\begin{tabular}{ccc}
 $p^2$ & A & B     \\ \hline  
{\small -2.66667}  & {\small -8.45038098750(31)} & {\small -8.45038} \\
{\small -1.77778}  & {\small -8.28747181582(32)} & {\small -8.28747} \\
{\small -1.18519}  & {\small -8.18480592994(33)} & {\small -8.18481} \\
{\small -0.790124} & {\small -8.11877664655(31)} & {\small -8.11878} \\
{\small -0.526749} & {\small -8.07577187774(28)} & {\small -8.07577} \\
{\small -0.351166} & {\small -8.04753658726(33)} & {\small -8.04754} \\
{\small -0.234111} & {\small -8.02890161217(30)} & {\small -8.02890} \\
{\small -0.156074} & {\small -8.01656069505(26)} & {\small -8.01656} \\
{\small -0.104049} & {\small -8.00836962582(21)} & {\small -8.00837} \\
{\small -0.069366} & {\small -8.00292496915(41)} & {\small -8.00292} \\
{\small -0.046244} & {\small -7.99930227781(36)} & {\small -7.99930} \\
{\small -0.030829} & {\small -7.99689023488(21)} & {\small -7.99689} \\
{\small -0.020553} & {\small -7.99528370180(10)} & {\small -7.99528} \\
{\small -0.013702} & {\small -7.99421324511(10)} & {\small -7.99421} \\
{\small -0.009135} & {\small -7.99349993358(10)} & {\small -7.99350} \\
{\small -0.006090} & {\small -7.99302446227(10)} & {\small -7.99302} \\
\hline
\end{tabular}
\end{center}
\caption{ Values of 
$T_{1234N}(\bar p^2;\bar m_1^2,\bar m_2^2,\bar m_3^2,\bar m_4^2)$
for small values of $\bar p^2=-p^2$. The masses are
$m_1=\bar m_2=3$, $m_2=\bar m_4=7$, $m_3=\bar m_3=5$, $m_4=\bar m_1=1$.
In column A are reported the results of the present method, with in brackets 
the error in the last digits, in column B are reported the results from 
column D of Table 1 of \cita{BBBB}, to whom barred notation refers.      }
\label{tab:tab2}
\end{table}
Our results are shown in column A of Table {\ref{tab:tab2}} and are in 
complete agreement with the results in column D of Table 1 of \cita{BBBB}, 
obtained from a one-dimensional integral representation and 
reported also here in column B for the reader's convenience. 
The results are purely real for small values of $|p^2|$, i.e. smaller 
than the smallest threshold value, which for the chosen values of the 
masses is the 3-body threshold of 
$T_{1234}^{(0)}(-p^2;m_4^2,m_1^2,0,0,\mu^2)$, 
corresponding to $p^2=-(m_1+0+0)^2=-9$.

Recently in \cita{Martin} a calculation with the same method has
provided the MDE and the expansion at $p^2=0$ for all the 2-loop
self-mass MI. 
Some numerical results are presented there for the function $U$ 
corresponding to $F_4^{(0)}$, that we have verified up to the 9 
digits reported in \cita{Martin}. 
The comparison is by no means simple, due to the different definition 
of the pole-term in the 4-denominator function in \cita{Martin}.
The finite terms of $U$ and $F_4$ differ and the term proportional to 
\( (n-4) \) of the Laurent-expansion of the 1-loop self-mass MI, 
that we take from \cita{JKV}, is necessary to account 
 for that difference.


\section{Conclusions.} 

We presented the results of the analytical expansion around $p^2=0$ 
and of the numerical calculation of the MI 
related to the general massive 2-loop 4-denominator self-mass diagram, 
using the 4-th order Runge-Kutta advancing solution method applied to 
the MDE satisfied by the MI. 

 An approximation method, suggested in \cita{Martin}, was properly developed
 here, allowing for precise numerical evaluation of the MI's even in the
 {\it surrounding-domains} of the {\it special points}.

 The relative precision reached by the developed numerical program
 is of the order of 10$^{-11}$--10$^{-12}$ for all points, but the 
 {\it surrounding-domain} of the 2-body threshold, where the relative
 precision is `only' of the order of 10$^{-8}$--10$^{-9}$.

 It is to be emphasized that the method requires only the knowledge 
 of Master Differential Equations (MDE's) satisfied by the MI's.
  The initial conditions
 follow from requirement of the regularity of MI at $p^2=0$ and MDE,
  while the form of the approximants used in the  
{\it surrounding-domains} of the {\it special points} follows from
 the MDE only.

{\bf Acknowledgments.}
 Henryk Czy\.z and Agnieszka Grzeli{\'n}ska are grateful to the Bologna Section
 of INFN and to the Department of Physics of the Bologna University
 for support and kind hospitality.

\vskip 2 truecm 

\def\NP{{\sl Nucl. Phys.}} 
\def\PL{{\sl Phys. Lett.}} 
\def\PR{{\sl Phys. Rev.}} 
\def\PRL{{\sl Phys. Rev. Lett.}} 
\def\NC{{\sl Nuovo Cim.}}
\def\APP{{\sl Acta Phys. Pol.}}
\def\ZP{{\sl Z. Phys.}}
\def\MPL{{\sl Mod. Phys. Lett.}} 
\def\EPJ{{\sl Eur. Phys. J.}} 
\def\IJMP{{\sl Int. J. Mod. Phys.}} 
\def\CPC{{\sl Comput. Phys. Commun.}}

\vfill \eject 
\vfill \eject 


\vfill \eject 


\end{document}